\documentclass[twocolumn,english,prl,notitlepage,nofootinbib,floatfix,longbibliography]{revtex4-1}
\usepackage[T1]{fontenc}
\usepackage[utf8]{inputenc}
\usepackage{refstyle}
\usepackage{amsmath}
\usepackage{bm} 
\usepackage{amssymb}
\usepackage{graphicx}
\usepackage[pdftex, hidelinks]{hyperref}
\usepackage[usenames,dvipsnames]{xcolor}
\usepackage{bbm}
\usepackage{tikz}
\usepackage{xspace}
\usepackage{bm}
\usepackage{booktabs}
\usepackage{babel}
\usepackage[separate-uncertainty=true]{siunitx}
\usepackage[makeroom]{cancel}
\usepackage{ifdraft} 

\usepackage[normalem]{ulem} 
\usepackage{comment}  
\usepackage{braket}  
\usepackage{float} 

\DeclareMathAlphabet\mathbfcal{OMS}{cmsy}{b}{n}

\makeatletter

\def\@fnsymbol#1{\ensuremath{\ifcase#1\or *\or *,\dagger\or \ddagger\or
   \mathsection\or \mathparagraph\or \|\or *\or \dagger\dagger
   \or \ddagger\ddagger \else\@ctrerr\fi}}

\AtBeginDocument{}
\RS@ifundefined{subsecref}
  {\newref{subsec}{name = \RSsectxt}}
  {}
\RS@ifundefined{thmref}
  {\def\RSthmtxt{theorem~}\newref{thm}{name = \RSthmtxt}}
  {}
\RS@ifundefined{lemref}
  {\def\RSlemtxt{lemma~}\newref{lem}{name = \RSlemtxt}}
  {}

\makeatother

\newcommand{\sref}[2]{\hyperref[#1]{\ref*{#1}(#2)}}

\newcommand\tauR{$\SI{40}{\micro\second}$ }
\newcommand\readpulse{$\SI{200}{\micro\second}$}
\newcommand\writepulse{$\SI{31}{\micro\second}$} 
\newcommand\opticalpumpduration{$\SI{350}{\micro\second}$ }

\newcommand\cellcavityFinesse{$\mathcal{F}\approx 13$}
\newcommand\CSparam{$\mathcal{R}$}
\newcommand\tauD{$\tau_\text{D}$}
\newcommand\magicdetuning{$\Delta_{4'} = 924$\,MHz}
\newcommand\zeemansplitting{$\nu_\text{L} = 2.4$\,MHz}

\newcommand\ttime{$T_2 = 2$ ms}

\newcommand\memorytimeshortnotation{$0.68\pm 0.08~\SI{}{\milli\second}$} 
\newcommand\memorytimeCSpar{$\tau_{\text{NC}}^{\mathcal R} = (0.68\pm 0.08) ~\SI{}{\milli\second}$}

\newcommand\timeviolatingbell
{$\tau_{\text{BI}} = (0.15 \pm 0.03) ~\SI{}{\milli\second}$}

\newcommand\memorytimeretrieval{$\tau_{\eta_\text{R}} = 0.89 ^{+ 0.49}_{- 0.23}~\SI{}{\milli\second}$}

\newcommand\BestCondAutoCorrResult{$g^{(2)}_{\text{RR|W=1}} = 0.20 \pm 0.07$}

\newcommand\gRRW{$g^{(2)}_{\text{RR|W=1}}$}

\newcommand\gWRcrosscorr{$g^{(2)}_{\text{WR}}$}

\begin{document}

\title{
Room-temperature single-photon source with 
near-millisecond
built-in memory
}

\author{Karsten B. Dideriksen}
\affiliation{Niels Bohr Institute, University of Copenhagen, Copenhagen, Denmark}
\author{Rebecca Schmieg}
\affiliation{Niels Bohr Institute, University of Copenhagen, Copenhagen, Denmark}
\author{Michael Zugenmaier}
\affiliation{Niels Bohr Institute, University of Copenhagen, Copenhagen, Denmark}

\author{Eugene S. Polzik}
\altaffiliation[Corresponding author: ]{polzik@nbi.ku.dk}
\affiliation{Niels Bohr Institute, University of Copenhagen, Copenhagen, Denmark}

\begin{abstract}
Non-classical photon sources are a crucial resource for distributed quantum networks.  Photons generated from matter systems with memory capability are particularly promising, as they can be integrated into a network where each source is used on-demand. Among all kinds of solid state and atomic quantum memories, room-temperature atomic vapours are especially attractive due to their robustness and potential scalability. To-date room-temperature photon sources have been limited either in their memory time or the purity of the photonic state. Here we demonstrate a single-photon source based on room-temperature memory. Following heralded loading of the memory, a single photon is retrieved from it after a variable storage time. The single-photon character of the retrieved field is validated by the strong suppression of the two-photon component with antibunching as low as \BestCondAutoCorrResult. Non-classical correlations between the heralding and the retrieved photons are maintained for up to \memorytimeCSpar, more than two orders of magnitude longer than previously demonstrated with other room-temperature systems.  Correlations  sufficient for violating Bell inequalities exist for up to \timeviolatingbell.
\end{abstract}

\maketitle

\section*{Introduction}
Remarkable progress has been achieved with deterministic solid-state single-photon sources 
\cite{Aharonovich2016,Awschalom2018,Wan2020,Ding2016,Kirsanske2017,Laplane2017,Kutluer2017}.
However, these sources require cryogenic temperatures to allow efficient photon interference 
\cite{Aharonovich2016,Awschalom2018,Lukishova2019}.
Ultracold atoms, which has been another system of choice 
\cite{Mucke2013,Radnaev2010,Corzo2019, Yang2016, Bimbard2014},
offer excellent performance at the expense of complexity of the experimental apparatus. In comparison, room-temperature atomic systems 
have attracted a lot of attention due to their potential scalability, robustness,
natural compatibility with atomic memories 
and favorable duty cycle \cite{Shu2016,Eisaman2005,Ripka2018,Finkelstein2018,Kaczmarek2018}.
The pioneering DLCZ proposal \cite{Duan2001} provided the route towards using an ensemble of atoms to combine single-photon generation and storage in the same system, offering experimental simplicity while enabling quantum information processing \cite{Nunn2013} and quantum communication \cite{Sangouard2011} schemes.
Envisioning quantum networks on a continental scale calls for quantum memories with storage time comparable to photon time-of-flight between the parties, i.e. in the millisecond regime.
The main challenge with room-temperature atomic ensembles is their thermal motion.
Introducing a buffer gas which slows down the atomic motion, allows for photon memory times of a few microseconds \cite{Dou2018,Pang2020, Bashkansky2012,Namazi2017,Reim2011,Hosseini2011}. 

Here we demonstrate an ensemble-based deterministic room-temperature single-photon source exhibiting clear antibunching and a non-classical memory of \memorytimeshortnotation{}, two orders of magnitude longer than previously demonstrated in room-temperature atomic vapours \cite{Dou2018,Bashkansky2012} and one order longer than our previous result \cite{Zugenmaier2018}.
This has been achieved by combining, for the first time, three main ingredients, the principle of motional averaging \cite{Borregaard2016, Zugenmaier2018}, a spin-protecting coating on the walls of the atomic vapour cell \cite{Balabas2010}, and the  use of a Raman transition at the "magic detuning" for writing and retrieving the single photon. 

We create a single collective excitation of the atomic ensemble when a heralding "write" photon generated via spontaneous Raman scattering is detected (Fig.\ \ref{fig:expSetup} \textbf{a}). 
Usually, the Gaussian transverse profile of the excitation beam leads to in-homogeneous coupling to the atoms, and therefore the detection of the heralding photon corresponds only to a "snap-shot" of the atomic positions. Consecutive atomic motion changes these positions and renders the subsequent retrieval of the photon inefficient. To remedy the effect of atomic motion we use motional averaging to project the ensemble onto the symmetric Dicke state with equal weights for all atoms \cite{Dicke1954}. This is achieved by narrowband filter cavities (Fig.\ \ref{fig:expSetup} \textbf{c}), extending the duration of the detection mode of the heralding photon beyond the transverse transit time of atoms through the cell channel. As the atoms travel through the beam, the random delay from the filter cavity leads via motional averaging to washing out the "which-path" information of the photon and thus equalizing the contribution of all atoms to the single collective excitation.
The antirelaxation coating of the walls preserves the spin state of the atoms for thousands of collisions, extending the lifetime of the symmetric collective excitation.
Four-wave-mixing (FWM) noise has been identified to be the main limitation for room-temperature vapour schemes \cite{Michelberger2015, Walther2007, Zugenmaier2018}. 
Several strategies have been pursued to suppress this noise including ladder schemes \cite{Kaczmarek2018,Finkelstein2018}, cavity engineering \cite{Saunders2016}, absorption \cite{Thomas2019} or Raman absorption \cite{Romanov2016}. 
An idea that is also suitable for Raman transitions between Zeeman levels
is to use polarization selection rules \cite{Walther2007,Zhang2014}. 
However, as a result of interfering excitation paths, this generally suppresses the Raman transitions as shown in \cite{Vurgaftman2013}. 
Here, we turn this effect to our advantage, exploiting a magic detuning (Fig.\ \ref{fig:expSetup} \textbf{b}) to suppress only the undesired FWM transition by the destructive interference of Raman amplitudes via coupling to different excited states (see Methods).

\section*{Experimental setup}

\begin{figure}
    \centering
    \includegraphics[width=\columnwidth]{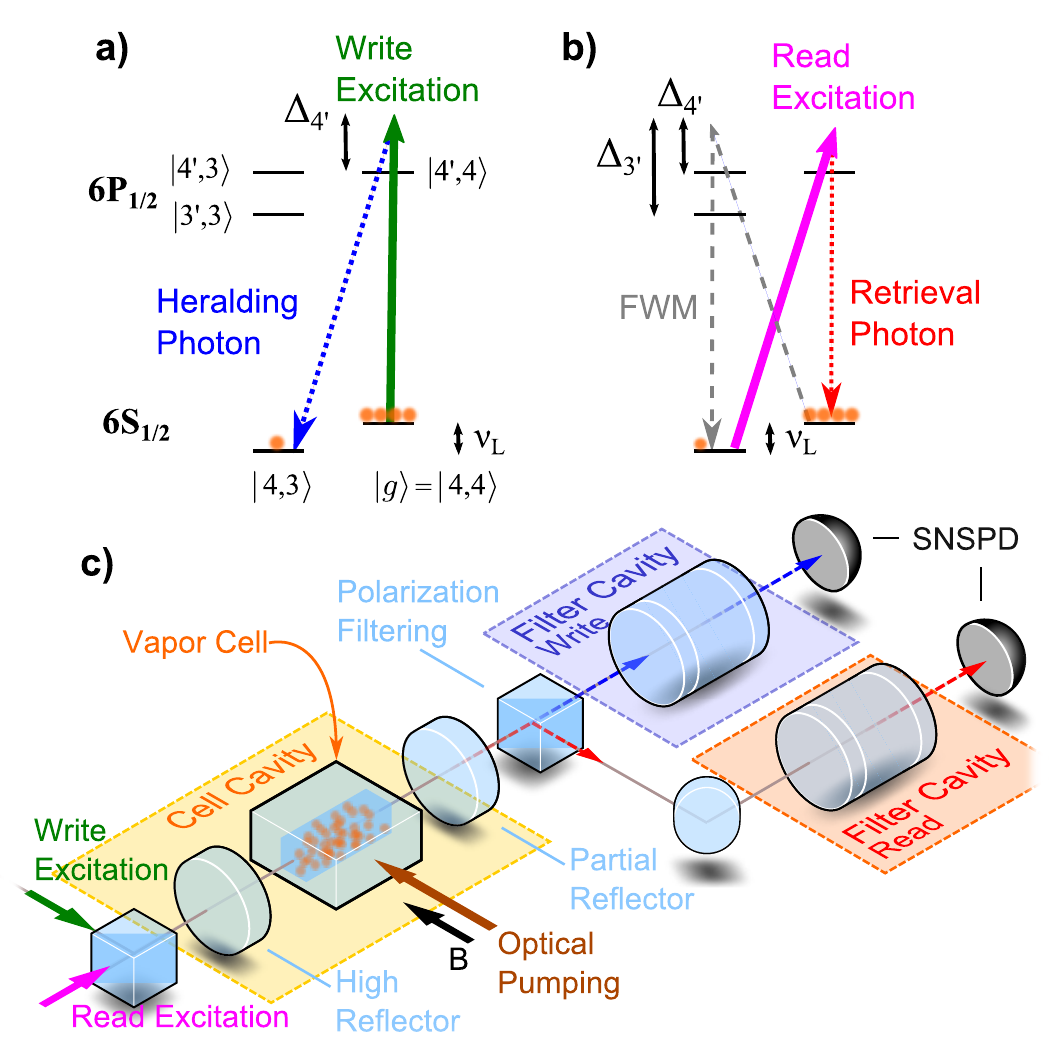}
    \caption{
    \textbf{a) Write excitation scheme:} 
    $\pi$-polarized, far-detuned excitation light creates an atomic excitation via a Raman scattering process. Only relevant atomic levels shown.
    \textbf{b) Read excitation scheme:}
    $\sigma$-polarized light used to retrieve stored excitation via Raman scattering and scattering desired deterministic single photon. Excess FWM noise is suppressed by choosing \magicdetuning .
    \textbf{c) Experimental setup:}
    Schematic of simplified experimental setup including  paths for write and read scattered photons through polarization and spectral filtering.    }
    \label{fig:expSetup}
\end{figure}%

In the experiment the atomic ensemble is a thermal caesium vapour with a cross-section of $\SI{300}{\micro\meter} \times \SI{300}{\micro\meter}$ and a length of $10\,$mm.
The $N$ atoms of the atomic ensemble are initially optically pumped into the coherent spin state  $\ket{g} = \ket{g_1 ... g_N}$, where $\ket{g_i} = \ket{F = 4, m_F = 4}$. Typically, we achieve an atomic polarisation of 99.2\%. Afterwards, a collective excitation is written into the atomic ensemble with low probability using a far-detuned, $\pi$-polarized write pulse (Fig.\ \ref{fig:expSetup} \textbf{a}). Upon detection of a scattered heralding single photon, the atomic ensemble is ideally projected onto a long-lived, symmetric Dicke state $\ket{s} = \sum_{i = 1}^{N} \frac{1}{\sqrt{N}} \ket{g_1 ... s_i...g_N}$ acting as the memory storage state with $\ket{s_i} = \ket{4,3}$.
The cell is subject to a magnetic field providing a frequency splitting of $\nu_\textrm{L} = \SI{2.4}{MHz}$ between the relevant Zeeman levels. To enhance light-atom interaction, the vapour cell  is placed in an asymmetric linear cavity (Fig.\ \ref{fig:expSetup} \textbf{c}) with finesse \cellcavityFinesse, a compromise between interaction enhancement and photon output coupling.
The orthogonal polarization of the heralding photon with respect to the excitation light and the relative detuning by one Larmor frequency $\nu_\textrm{L}$ facilitates filtering of the heralding  photon from the $10^7$ excitation photons in the same spatial mode by polarization filtering optics and subsequent spectral filtering with  narrowband filter cavities (Fig.\ \ref{fig:expSetup} \textbf{c}). 
The filter cavities simultaneously serve the purpose of motional averaging by adding random delays to the scattered photons \cite{Borregaard2016}, selecting the symmetric Dicke state.

After a variable delay \tauD, a $\sigma$-polarized read pulse retrieves the stored collective excitation coherently in form of a deterministic "retrieval" single photon (Fig.~\ref{fig:expSetup}~\textbf{b}). 
The filtered heralding and retrieval single photons are detected using two superconducting nanowire single-photon detectors (SNSPD).

\begin{figure}[H]
    \centering
    \includegraphics[trim=35 7 120 80, clip, width=\columnwidth]{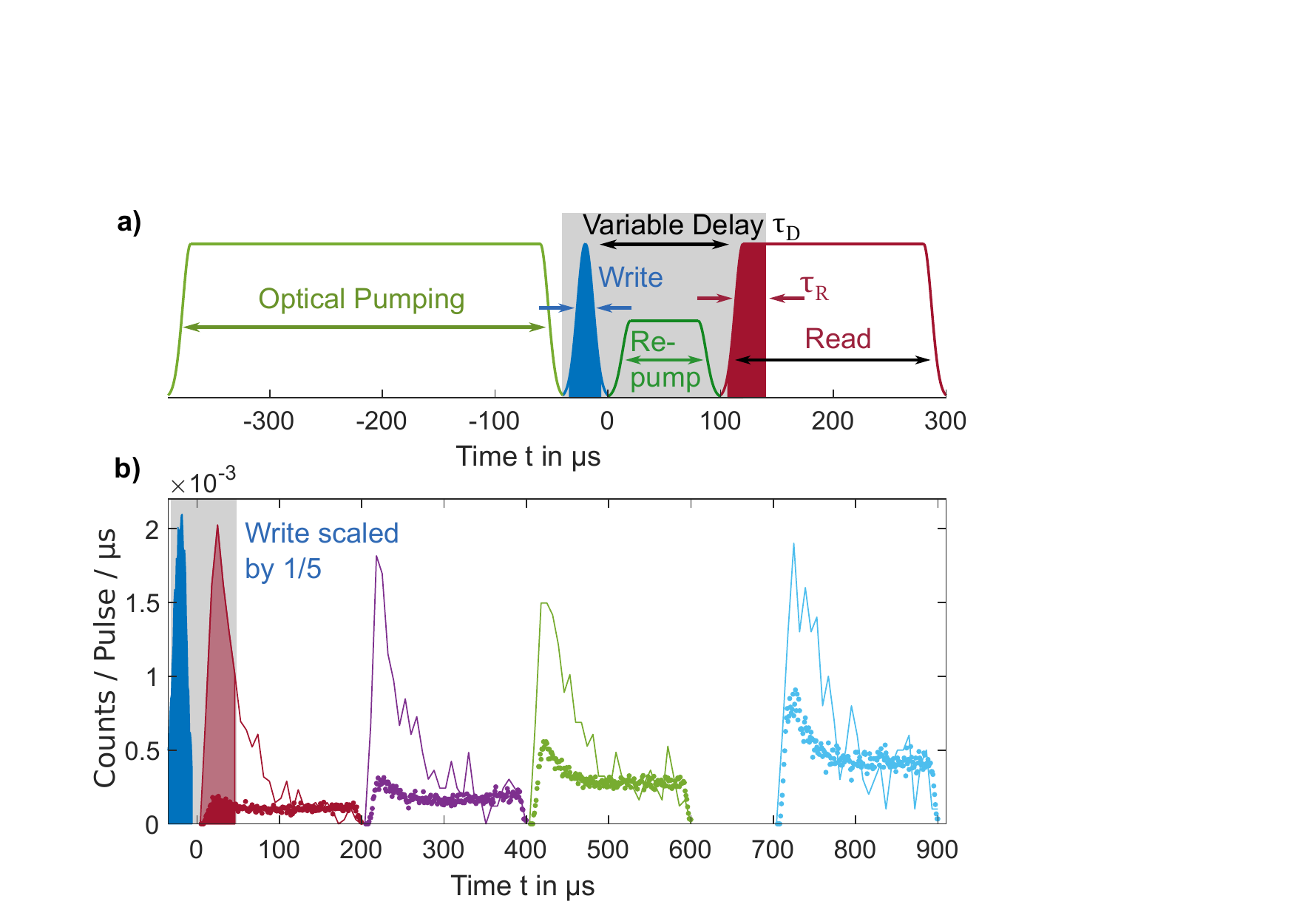}
    \caption{ \textbf{a) Experimental pulse sequence:} Illustration of smoothened write, read and optical pumping pulses, variable delay between write and read, and optional repump for delayed readout. The solid areas represent integration windows used in the analysis. 
    \textbf{b) Temporal shape of detection events:}
    Detected counts during write (\writepulse, scaled 1/5, blue) and read pulses (\readpulse, delayed 10 to  $\SI{710}{\micro\second}$) with noise level (dotted, $\SI{1}{\micro\second}$ binning) and detection events conditioned on the heralding write (lines, $\SI{7}{\micro\second}$ binning).
    }
    \label{fig:histograms_cond_uncond_readpulses}
\end{figure}%

The excitation light and single photons propagate along the cell axis orthogonal to the quantization axis defined by the optical pumping and magnetic field (Fig.~\ref{fig:expSetup}~\textbf{c}). The write excitation light in $\pi$-polarization (solid green arrow in Fig.~\ref{fig:expSetup}~\textbf{a}) generates the write photon (dashed blue) in the polarization mode orthogonal to the quantization axis. In the read process the latter polarization mode is used by the linearly-polarized excitation light. 
Beside the desired $\sigma_+$-polarized component (solid pink arrow in Fig.~\ref{fig:expSetup}~\textbf{b}) this linearly-polarized mode contains also the undesired $\sigma_-$-polarized component (dashed diagonal line in Fig.~\ref{fig:expSetup}~\textbf{b}) which drives the FWM process contaminating the stored excitation and adds noise to the retrieved light.
By choosing the magic detuning \magicdetuning ~for the read we  effectively suppress this FWM process that turns out to be critical for the purity of the generated state.

The experimental sequence consists of two main parts, an optical pumping and locking window for all cavities, and a window containing the experimental write-read pulse sequence. The latter (Fig.\ \ref{fig:histograms_cond_uncond_readpulses} \textbf{a}) contains a \opticalpumpduration optical pumping pulse for state initialization, a $\SI{40}{\micro\second}$ write pulse, a variable delay \tauD, and a  \readpulse~read pulse. All pulses are turned on and off smoothly to prevent high frequency harmonics which can falsely excite the memory state.
The sequence of write, read and optical pumping pulses is repeated up to 75 times, depending on \tauD ~before re-locking of the cavities becomes necessary. For delay times of $\SI{100}{\micro\second}$ and longer, an additional repump pulse is used to counteract birefringence effects in the cell cavity due to the atomic polarization decay.

The retrieved light follows an exponentially decaying temporal envelope (Fig.\ \ref{fig:histograms_cond_uncond_readpulses} \textbf{b}). 
One component of the atomic noise follows the same envelope while the second component grows linearly during the delay and the pulse duration (the noise origin is discussed below). Hence, the signal-to-noise ratio (SNR) depends on the time window chosen in post-processing. We find that truncating the write window to \writepulse\ out of the \SI{40}{\micro\second} pulse duration and the read window at $\tau_\textrm{R} =$ \tauR offers a good trade-off between SNR and retrieval efficiency, see Supplementary Information (SI).

\begin{figure}[H]
    \centering
    \includegraphics[trim=25 0 120 0, clip,width=\columnwidth]{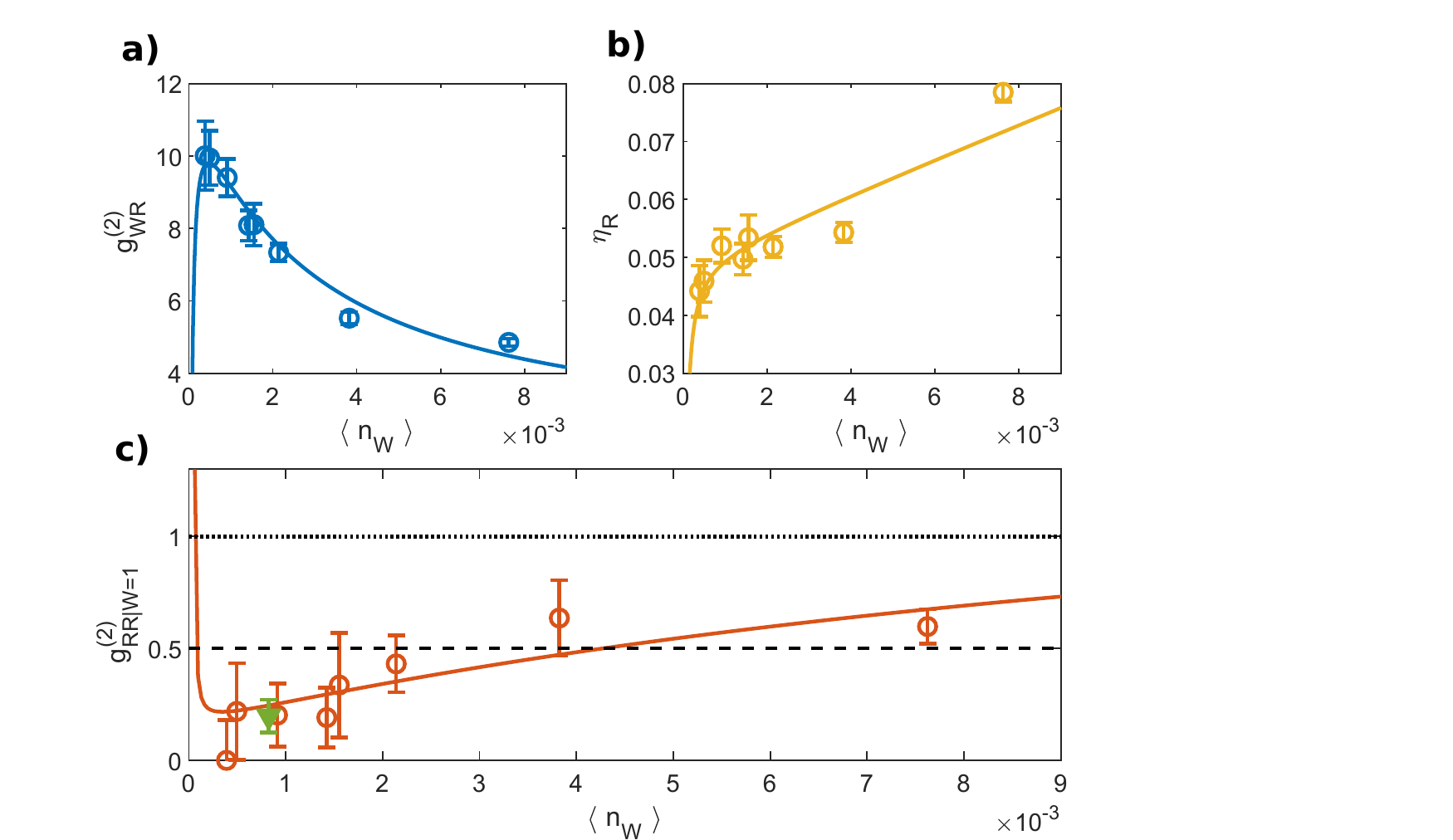}
    \caption{
    \textbf{a)} 
    2$^{\text{nd}}$-order cross-correlation function \gWRcrosscorr{},
    \textbf{b)} retrieval efficiency $\eta_\text{R}$ and \textbf{c)} 2$^{\text{nd}}$-order conditional auto-correlation function  \gRRW{} of the retrieved light field. \gRRW $=1$ (dotted line) is the classical limit and \gRRW $=0.5$ (dashed line) is the two-photon Fock state auto-correlation value.
    All functions are plotted against the mean number of detected write counts $\braket{n_\text{W}}$.
    Shown are measured data (circles) and the model (full lines). To improve the experimental uncertainty on \gRRW{} we combine points for $\braket{n_\text{W}}<2\cdot 10^{-3}$ (green triangle).  Error bars represent one standard deviation.}
    \label{fig:cross_cond_correlation_and_RetEffvsp0}
\end{figure}%

\section*{Photon correlations}
The conditional generation of a single excitation in the atomic memory is characterized by the non-classical cross-correlations between the single photons scattered during the write and consecutive read pulses.
The relevant 2$^{\text{nd}}$-order cross-correlation is given by 
\gWRcrosscorr{} = $\left< n_\text{W}n_\text{R}\right>/(\left<n_\text{W}\right>\left<n_\text{R}\right>)$
where ${n_\text{W}}$ $({n_\text{R}})$ is the number of detection events during the write (read) process.

A long temporal shape of the retrieved light in the tens of microseconds range provides an advantage for the characterization of the photon source.
Under those conditions the SNSPD, for which the dead time is less than 50 ns, works as a photon-number-resolving detector. This capability allows for accurate accounting of multi-photon events which would otherwise compromise the accuracy of the measurement of correlation functions. 

In the absence of losses and extra noise the joint state of the write photon and the memory is of the two-mode squeezer type:
\begin{multline}\ket{\Psi_\textrm{uncond.}} = \sqrt{1-p_0}\Big(\ket{0}_\textrm{W}\ket{0}_\textrm{A} + \sqrt{p_0}\ket{1}_\textrm{W}\ket{1}_\textrm{A} + \\ \left. p_0\ket{2}_\textrm{W}\ket{2}_\textrm{A} + \mathcal{O}\left(p_0^{3/2}\right)\right)
\end{multline}
where $p_0$ is the probability of creating one or more excited pairs. $\ket{n}_\textrm{W}$ ($\ket{n}_\textrm{A}$) refers to $n$ excitations of the write scattered field (symmetric excitations in the atomic ensemble). 
Thus, the multiple-pair excitation probability $p_0$ has to be kept low enough to avoid falsely heralding the single-pair state due to limited detection of the heralding field (e.g. propagation losses). 
We can directly relate the excitation probability to the mean number of excitations $\left<n_\textrm{exc}\right>$ via $p_0 = \left<n_\textrm{exc}\right>/(1+\left<n_\textrm{exc}\right>) $. For low number of excitations and neglecting noise, this gives $p_0 \approx \left<n_\textrm{W}\right>/\eta_{\textrm{X}}$, which is the mean number of detected write counts 
$\left<n_\textrm{W}\right>$ scaled with the
write detection efficiency $\eta_{\textrm{X}}=2.9\%$
that includes the outcoupling from the cell cavity, propagation efficiencies through the filter setup and the quantum efficiency of the detector, see SI. 

With the decreased write pulse energy and thus $\left<n_\textrm{W}\right>$, \gWRcrosscorr{} grows as seen in Fig.~\ref{fig:cross_cond_correlation_and_RetEffvsp0} \textbf{a} as a low multi-pair generation probability is crucial for a high cross-correlation between the write and read fields \cite{Sangouard2011}.
When $n_\textrm{W}$ would be decreased even further, we expect the detection events of the write field to be dominated by background noise limiting the correlations.
The high value of \gWRcrosscorr{}$~\approx 10$ obtained for low $n_\textrm{W}$ testifies to the high heralding efficiency of the excitation storage in the memory and its consecutive readout.

The retrieval efficiency $\eta_\text{R}(\tau_\text{R}) = \braket{n_\text{R|W=1}(\tau_\text{R})}- \braket{n_
{\textrm{noise}}(\tau_\text{R})} $ is defined as the difference between the mean number of counts conditioned on a single heralding write count, and the mean number of detected noise read counts in the absence of a write pulse (SI). 
As $\braket{n_\textrm{W}}$ grows
(Fig.~\ref{fig:cross_cond_correlation_and_RetEffvsp0} \textbf{b}), $\eta_\textrm{R}$ first grows rapidly as the write dark counts become negligible, and then continues to grow slower
as the heralded state acquires an increasing contribution of multiple stored excitations.

To gain more insight, we model the system as a two-mode squeezed state with uncorrelated noise using probability generating functions (SI).
The model yields the mean detected count rates, the cross-correlation, the retrieval efficiency and the conditional auto-correlation. The atomic noise contributions used in the model are found experimentally from the spectral scans of the filter cavities with and without sending a write pulse (SI).
The only free fit parameters remaining in the model are the detection and intrinsic retrieval efficiencies. 
These are determined by simultaneously fitting to \gWRcrosscorr{}, $\eta_\textrm{R}$ and $\left<n_\textrm{R}\right>$. We observe good agreement of the experimental data with the fitted model as seen from  Fig.~\ref{fig:cross_cond_correlation_and_RetEffvsp0}.
From the fit parameter we estimate the intrinsic retrieval efficiency, i.e.\ the efficiency of retrieving one excitation from the symmetric atomic mode into the cell cavity mode, to be $\eta^*_\text{R} = (70\pm8)\%$ for $\tau_\text{R}=$ \tauR (see Methods).

Next, we demonstrate that the memory indeed stores a single excitation which can be deterministically retrieved on-demand as a single photon.
Towards this end, we measure the conditional auto-correlation function and verify the sub-Poissonian character of the retrieved field, for which \gRRW{} < 1. Fig.~\ref{fig:cross_cond_correlation_and_RetEffvsp0} \textbf{c} shows \gRRW{} as a function of $\braket{n_\text{W}}$ 
which in the present case of number-resolving detection is defined as \gRRW{}
$ = \left< n_\textrm{R|W=1} (n_\textrm{R|W=1}-1)\right>/\left<n_\textrm{R|W=1}\right>^2$ where $n_\textrm{R|W=1}$ is the number of read detection events in each sequence with a preceding heralding write detection event. 
We observe good agreement between the experimental data and the model.

To improve the precision of \gRRW{} 
we combine datasets for $\braket{n_\text{W}}<2\cdot 10^{-3}$. According to the model, the read field is found to weakly depend on $p_0$ in this range. Under those conditions, the write-read sequence has been repeated $3\cdot 10^7$ times over 32 hours of run time. 
In 1296 cases, we successfully created and retrieved a symmetric collective excitation.
A double read detection event has been registered in $7$ sequences resulting in \BestCondAutoCorrResult, a clear indication of the sub-Poissonian nature of the conditional read field for a read delay of \SI{10}{\micro\second}. Furthermore, there is an appreciable margin of more than four standard deviations to the two-photon Fock state auto-correlation $g^{(2)}_\textrm{n=2,n=2}=0.5$, which indicates good fidelity of the single-photon state.

\section*{Delayed readout \& memory time}
The quantum memory capabilities of the system are mapped out 
by varying the delay time between write and read pulses from 10 to 1010 $\SI{}{\micro\second}$.
In Fig.\ \ref{fig:histograms_cond_uncond_readpulses}\textbf{b} we have included the histograms for conditional read counts and unconditional read noise for various delays.
While in the dark, the atoms decay into the storage state primarily due to wall collisions. This leads to an increase of the readout noise with $\tau_\textrm{D}$. Atomic decay compromises the readout in two ways, as incoherently transferred atoms in $\ket{4,3}$ contribute to both symmetric and asymmetric modes of the ensemble:
1) The high readout rate at the beginning of the read pulse originates from the symmetric mode which is read out in the same temporal shape as the desired stored excitation. This constitutes approximately half of the noise at the beginning of the read pulse.
2) The asymmetric modes are read out inefficiently, leading to a count rate that slowly increases over time as the population in $\ket{4,3}$ grows. The noise from asymmetric modes dominates the readout at the end of the read pulse.
Both noise contributions grow approximately linearly with $\tau_\textrm{D}$ degrading the SNR between conditional and unconditional readout.

\begin{figure}[H]
    \centering
    \includegraphics[trim= 5 0 20 10, clip,width=\columnwidth]{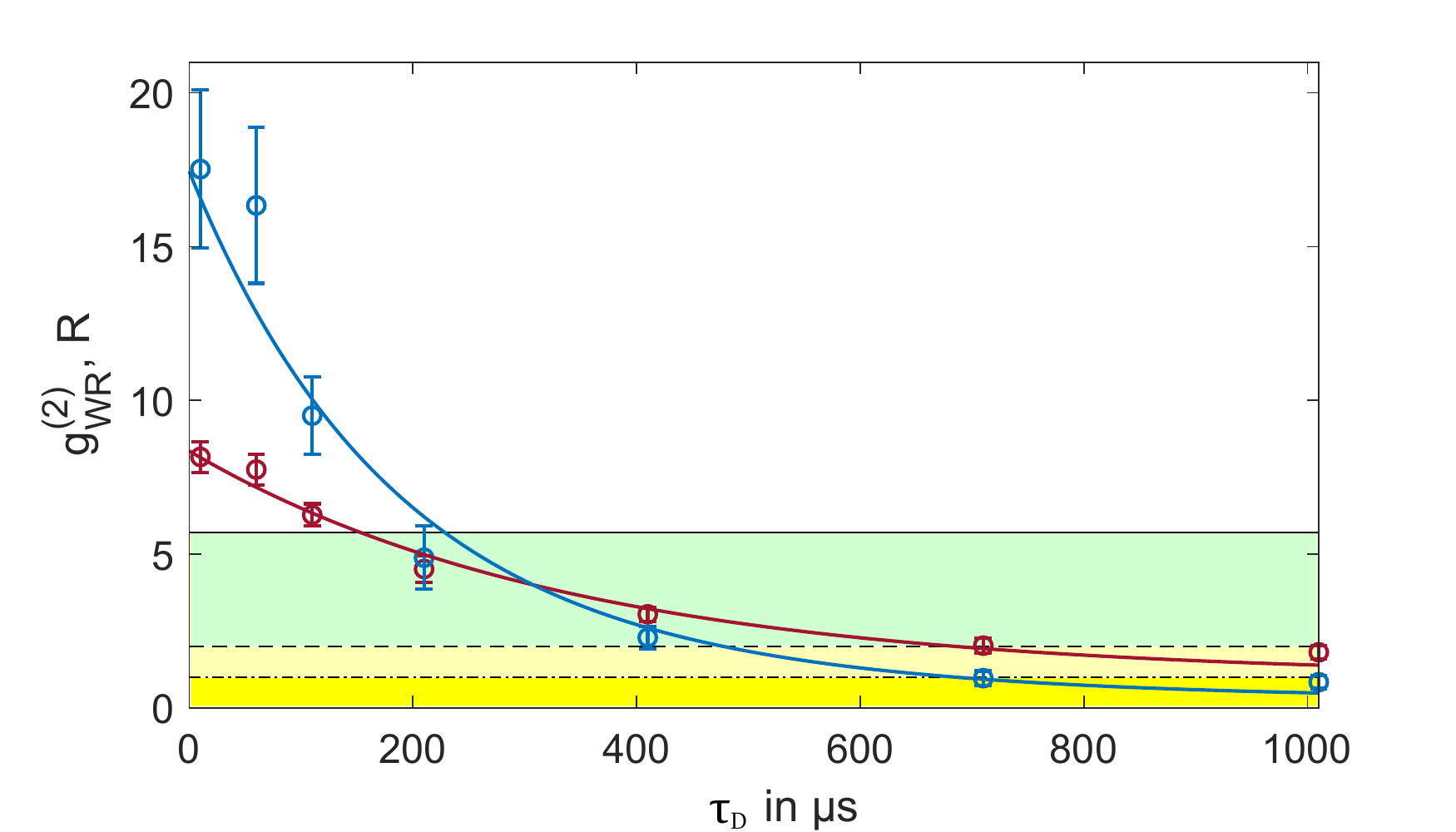}
    \caption{
    \textbf{ Photon correlations for delayed readout:}
    Shown are \gWRcrosscorr{} (red) and the Cauchy-Schwarz parameter \CSparam ~ (blue) versus various read pulse delays \tauD ~for an integrated read pulse duration of \tauR     together with the fit to \gWRcrosscorr{} (red line) and the resulting  \CSparam ~ (blue line). The black line marks the Bell-inequality limit \gWRcrosscorr{} $ \geq 5.7$ and the dashed line marks the typical non-classicallity signature \gWRcrosscorr{} $ > 2$. The dash-dotted line is the formal non-classicallity criterion \CSparam\ $ > 1$. Error bars represent one standard deviation.}
    \label{fig:Cross_Rparam_vs_delay}
\end{figure}%

Important characteristics of the on-demand single photon source enabled by the quantum memory are the 2$^{\text{nd}}$-order cross-correlation function \gWRcrosscorr{} and the Cauchy-Schwarz parameter \CSparam\ = ${\left(g^{(2)}_{\text{WR}}\right)^2}/\left({g^{(2)}_{\text{RR}}\,g^{(2)}_{\text{WW}}}\right)$ versus delay time \tauD~\cite{Clauser1974}. The latter can be used to quantify the non-classicality of correlations between write and read intensities.
In Fig.\ \ref{fig:Cross_Rparam_vs_delay} the respective values for \gWRcrosscorr{}(\tauD) and \CSparam(\tauD) are shown, along with the exponential fit following \gWRcrosscorr{}(\tauD) $= B\cdot\exp(-\text{\tauD}/\tau_\text{M}) + 1 $. 
$g^{(2)}_{\text{WW}}$ is independent of \tauD\ and because noise dominates the unconditional readout, the dependency for $g^{(2)}_{\text{RR}}$ is marginal 
which is what we observe.
We therefore use averaged values for $g^{(2)}_{\text{WW}}$ and $g^{(2)}_{\text{RR}}$ together with the fit results of \gWRcrosscorr{}(\tauD) to plot the above expression for \CSparam(\tauD).
From \CSparam\ we define the memory time as the time beyond which write and read light fields are no longer non-classically correlated, i.e.\ not fulfilling \CSparam\ $ > 1$ (Fig.\ \ref{fig:Cross_Rparam_vs_delay}, dash-dotted line). 
We use \CSparam\ as a formal non-classicality bound instead of the typical signature \gWRcrosscorr{} $ > 2$ \cite{Sangouard2011}.
The corresponding memory time is \memorytimeCSpar. The limit for violating the Bell inequality is given by \gWRcrosscorr{}$ \ge$  5.7 \cite{Wallucks2020}. From the fit in Fig.\ \ref{fig:Cross_Rparam_vs_delay} this holds for \timeviolatingbell.

The non-classical memory time of the atomic ensemble is limited by noise from the atomic decay. However, the retrieval efficiency $\eta_\textrm{R}$ is noise free due to its definition (SI), which allows us to determine the $1/e$-lifetime of the collective excitation in the memory, amounting to \memorytimeretrieval.
The collective excitation lifetime is expected to be limited to half of the transverse macroscopic spin amplitude decay time, separately measured to be \ttime{} (Methods) and to be dominated by spin relaxation due to wall collisions.

\section*{Discussion}
Our results demonstrate the capability to herald, store and read out a single long-lived collective atomic excitation from a room-temperature atomic vapour.
We verify the single-photon nature of the retrieved light from observing strong photon antibunching.
High cross-correlation and near-millisecond storage time at room temperature enable applications in quantum networks, where the platform can be immediately used as a building block for entanglement generation over up to 200\,km. The feasibility of this was demonstrated in a recent proof-of-principle study verifying short-range entanglement of two warm atomic vapours after sub-microsecond storage through the DLCZ protocol \cite{Li2020}.
The technological simplicity of our system facilitates spatial multiplexing and thus quantum repeater as well as simulator applications.
Straightforward modifications, such as improved filling of the cell by the excitation beam and a larger cell cross section would further improve efficiency, fidelity and storage time to the timescale of seconds \cite{Katz2018}.
Besides the application as a source of narrowband single photons one can explore a larger phase space  by accumulating excitations.
Exciting applications may also arise from interfacing such quantum-state engineering with other platforms, such as cold atoms or mechanical oscillators.

\section{Methods}
\label{app:Methods}
\textbf{Light.} Excitation light pulses for write and read are derived from a narrowband home-built external cavity diode laser at 895 nm. It is locked via a beat-note lock with fixed detuning to the $F = 4 \rightarrow F' = 4$ transition of the $D_1$ line of caesium. The write and read locking and excitation frequencies are derived using two AOMs. 
\bigskip

\textbf{Vapour cell.} In our experiments we use a caesium vapour cell with an interaction volume of \SI{300}{\micro\meter} x \SI{300}{\micro\meter} x 10 mm, coated with a spin-preserving anti-relaxation coating (alkane). The cell cavity mode has a \SI{90}{\micro\meter} waist radius ($1/e^2$ intensity) at the cell center.
Using magneto-optical resonance spectroscopy \cite{Julsgaard2004}, the coherence time of the ground-state Zeeman levels was determined to be \ttime \, for an operational temperature of 43  $^\circ$C of the experiment and a Zeeman splitting of \zeemansplitting.\bigskip

\textbf{Data acquisition.} To compensate for drifts in the experimental setup while acquiring measurement data, sequences with and without write, as well as sequences with varying delay \tauD~are interleaved. Sequences without preceding write pulse are used to estimate noise levels in the readout.\bigskip

\textbf{Optical pumping.}
During the locking and optical pumping window in the experimental sequence, the coherent atomic spin state is prepared using two circularly polarized pump and repump beams. The pumping is parallel to the magnetic field orientation. The repump laser is locked onto the $F = 3 \rightarrow F' = 2,3$ crossover transition, while the pump laser is locked on the transition $F = 4 \rightarrow F' = 4$. We determine the atomic polarization of atoms in the $F = 4$ manifold (typically > 99.2) using pulsed magneto-optical resonance spectroscopy. This high polarization is achieved by optimized beam geometry and by turning off the repump laser first and keeping the pump laser turned on for a few microseconds longer.\bigskip

\textbf{Polarization and spectral filtering.}
The leakage contribution is minimized using a half wave plate and a quarter wave plate after the cell cavity to optimize the polarization orientation such that the polarization filtering using a Glan-Thompson polarizer reaches a suppression of $5\cdot10^{-5}$. 
Following this polarization filtering stage, the spectral filtering consisting of two cavities for each of the detection setups, provides around 60 dB suppression for both detection setups. \bigskip

\textbf{Magic detuning.}
FWM noise is due to read excitation light coupling to the state $\ket{4,4}$  and via a spontaneous Raman process creating excess excitations in $\ket{4,3}$.
The associated Raman-Rabi coupling is given by $R \propto g \Omega/\Delta$, where $\Delta$ is the detuning, 
 and $\Omega$ ($g$) is the coupling strength for the excitation field (scattered field), respectively. These coupling strengths include the Clebsch-Gordan coefficients for the corresponding transitions.
For the Raman transition coupled to multiple excited states $m$ we need to sum over their contributions to the coupling $R \propto \sum_{m}  g_{m} \Omega_{m}/\Delta_{m}$, where $\Delta_{m}$ is the detuning from the respective state. 
For Raman transitions with Clebsch-Gordan coefficients of opposite signs, there will be a detuning where the above sum vanishes. 
For caesium atoms and light on the $D_1$ line with a Raman transition between the states $\ket{4,4}$ and $\ket{4,3}$ via the excited states $\ket{4',3}$ and $\ket{3',3}$, the detuning where this transition is effectively suppressed lies outside the Doppler-broadened width.
Including the motion of the atoms we can follow the derivation for motional averaging from \cite{Borregaard2016}, adding the relevant excited states. 
This yields the expression for the coupling $R \propto \sum_{m} g_{m} \Omega_{m} w[(\Delta_{m} + i \gamma/2)/\Gamma_D]$ with the Faddeeva function $w[z]$, the natural linewidth $\gamma$ and the Doppler broadening $\Gamma_D$.
We find an optimal FWM suppression at a detuning of \magicdetuning.
\bigskip

\textbf{Intrinsic retrieval efficiency.}
From the correlation model (see SI) we find the fit parameter $\eta_Y = (6.0\pm0.2)\%$ which is the probability to have a detection event caused by retrieving one collective excitation in the symmetric atomic mode. Thus, it includes propagation losses, i.e.\ $\eta_Y = \eta_\text{d}\eta_\text{esc}\eta^*_\text{R}$. To estimate the intrinsic retrieval efficiency $\eta^*_\text{R}$ we correct for the losses from two parts of the setup: 
1) The efficiency $\eta_\textrm{esc}$ of a photon generated inside the cell cavity escaping out through the outcoupling mirror. It is found from the single-pass transmission through the vapour cell $T_\textrm{cell}$ and the reflectivity $R$ of the outcoupling mirror as $\eta_\textrm{esc} \approx T_\textrm{cell}\left( 1 - R\right) / \left( 1 - RT_\textrm{cell}^2\right) = (45\pm 2)\%$.
2) The detection efficiency of light in the retrieval photon mode after the cell cavity. We determine this using strongly attenuated laser light in the retrieval photon mode to be $\eta_\text{d}=(19\pm 2)\%$.
\bigskip

\textbf{Uncertainty estimation.}
The error bars for conditional auto-correlation functions and the cross-correlation functions, as well as the retrieval efficiency are calculated using Poissonian errors. An exception is the error bar on the \gRRW\ $=0$ point in Fig.\ \ref{fig:cross_cond_correlation_and_RetEffvsp0}\ \textbf{c} for the conditional auto-correlation where no conditional double read detection event was recorded. In this case, we assign an error bar of \gRRW\ equal to the value if one conditional double read detection event had been recorded.
\bigskip

\section*{Acknowledgments}
The authors acknowledge helpful discussions with Michal Parniak and support with the design of cavities by Ivan Galinskiy. The alkane coated cell has been fabricated by Michail Balabas.

This project has been supported by the European Research Council Advanced grant QUANTUM-N, the Villum Foundation, and John Templeton Foundation.

\section*{Author Contributions}
E.\,S.\,P.\ conceived and led the project. K.\,B.\,D., R.\,S. and M.\,Z. have contributed equally to the work, built the experiment, took the data and performed the analysis. All authors discussed the results and participated in the writing process of this manuscript.

\section*{Data Availability Statement}
Source data are available for this paper. All other data that support the plots within this paper and other findings of this study are available from the corresponding author upon reasonable request.

\section*{Competing interests}
The authors declare no competing interests.

\appendix
\clearpage

\setcounter{page}{1}
\renewcommand{\thepage}{SI~\arabic{page}}

\setcounter{figure}{0}
\renewcommand{\thefigure}{SI\arabic{figure}}

\setcounter{table}{0}
\renewcommand{\thetable}{SI\arabic{table}}

\setcounter{equation}{0}
\renewcommand{\theequation}{SI~\thesection.\arabic{equation}}

\clearpage
\onecolumngrid
\section*{Supplementary Information}

\section{Experimental setup}
\label{app:Setup}

\begin{figure}[H]
    \centering
    \includegraphics[trim= 530 0 0 200, clip, angle=0, width=160mm]{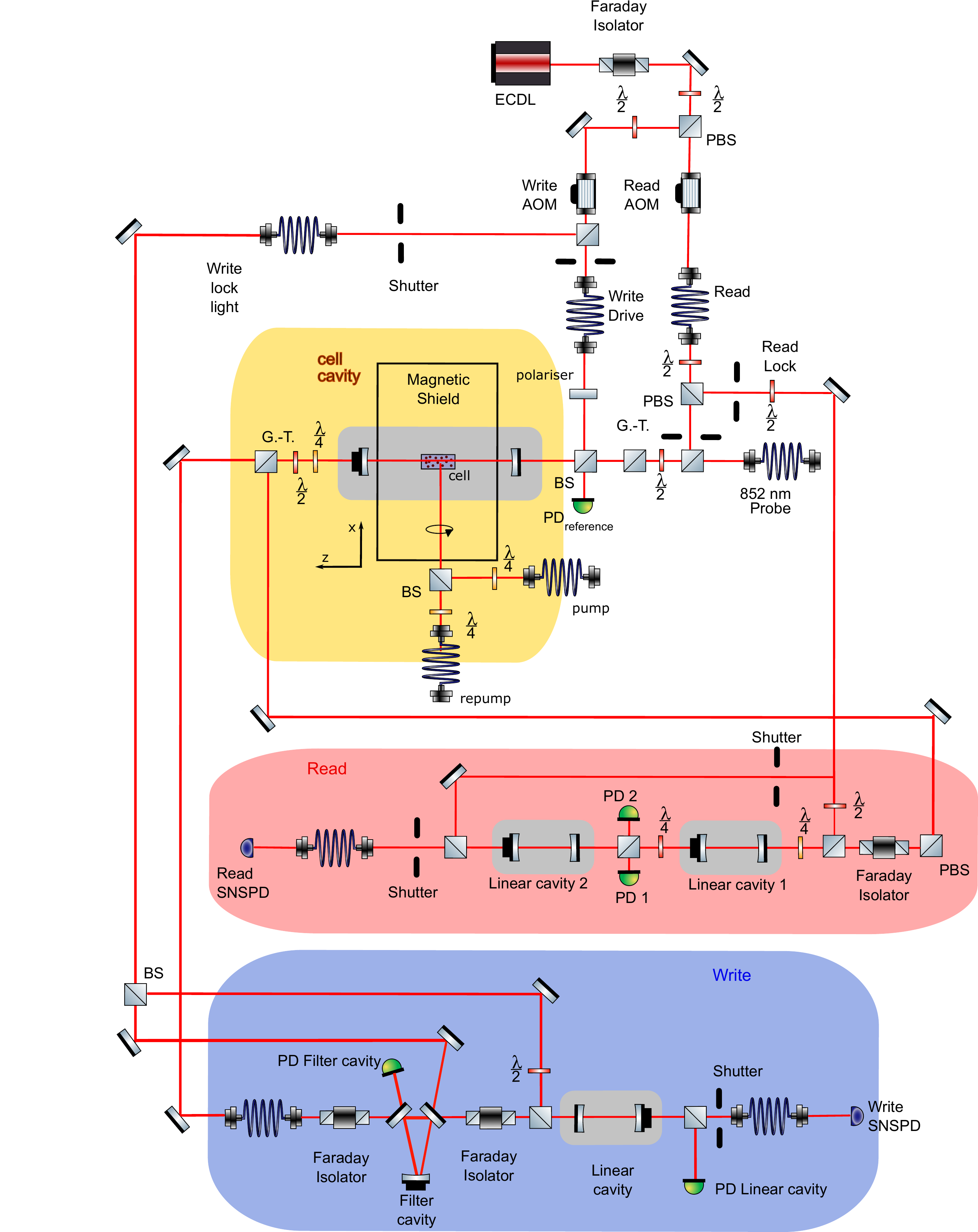}
    \caption{\textbf{Schematic of Setup: } Drawing of experimental setup including optical pumping, excitation and lock light paths, along with scattered photon paths through write and read filtering and detection setups. The photo detectors (PD) are used for locking. The Glan-Thompson polarisers are denoted G.-T. in the schematic. The write filtering setup consists of a triangular and a linear cavity, while for the read filtering two linear cavities are used.
    }
    \label{fig:CompleteSetup}
\end{figure}%

In Fig.\ \ref{fig:CompleteSetup} we present a detailed schematic of the experimental setup. 
It shows all paths including the excitation light paths, heralding and retrieval scattered photons and beam paths used for locking of the cavities. As can be seen in Fig.\ \ref{fig:CompleteSetup}, the high excitation light suppression is facilitated through two consecutive cavities in each of the detection setups.

The various cavities in the setup are stabilized on resonance using different methods. The linear and the triangular cavities in the write filtering and detection setup are locked using the transmission of light modulated around the scattered write photon frequency during the locking time window of the experimental sequence. 
The error signal from the transmission signal is fed back onto the respective cavity's piezo-actuated mirror.
For the two consecutive linear cavities employed in the read filtering and detection setup we use dithering of the piezo-actuated mirrors to derive an error signal based on the transmission signal.
The cell cavity is locked  via piezo dither locking using a frequency-stabilized laser at 852\,nm. Its frequency is adjusted such that we achieve simultaneous resonance of locking light (852 \,nm) and scattered photons (895 \,nm).
The cell cavity linewidth is an order of magnitude broader than the Larmor frequency, thus the excitation beams are also close to resonance.
Using this laser, the error signal for locking the cell cavity is generated using dithering of the piezo-actuated mirror.

\section{Spectrum of Write and Read scattered photons}
\label{app:FCavSpectrum}
\begin{figure}[H]
    \centering
    \includegraphics[width=140mm]{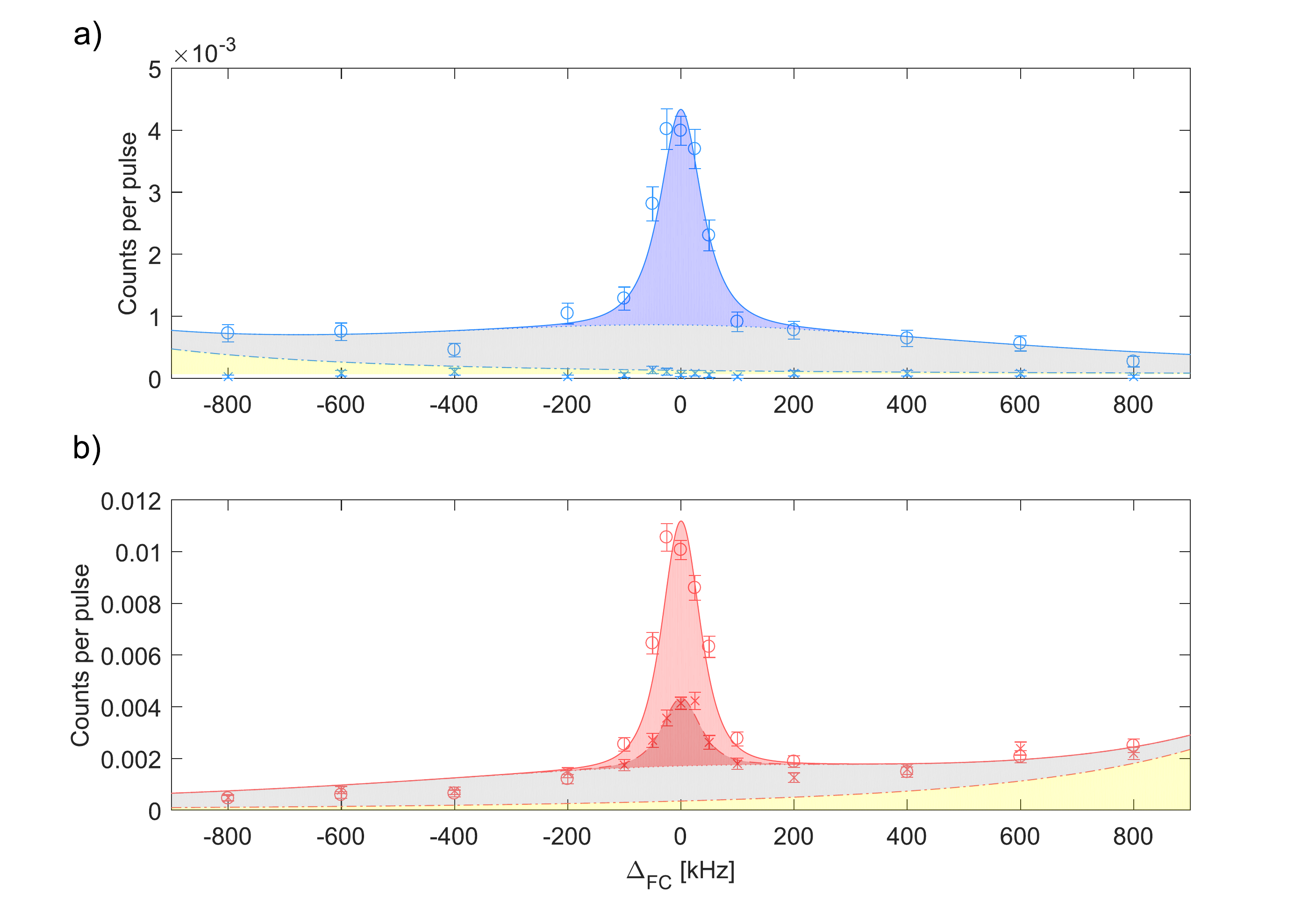}
    \caption{\textbf{Spectrum of scattered write and read photons: }
    Detected unconditional counts per pulse during \textbf{a) Write} and \textbf{b) Read} with $\tau_R=$\tauR and minimal read delay \tauD\ for different filter cavity detunings $\Delta_\text{FC}$ measured with write pulse (circles) and without sending a write pulse (crosses). The fits (lines) are shown with colored areas for narrowband (blue and red for write and read, respectively), broadband (gray) and leakage (yellow) contributions.
    }
    \label{fig:FCavScan}
\end{figure}

The spectrum of the write and read scattered photons is taken by detuning the frequency of the filter cavities by $\Delta_\text{FC}$ with respect to the photons' frequency (Fig.~\ref{fig:FCavScan}). 

In the present excitation scheme the heralding photon (anti-Stokes photon) is scattered   blue-shifted by the Larmor frequency \zeemansplitting\ compared to the write excitation light while the retrieved single photon is red-shifted by  $-\nu_\mathrm{L}$. For the write process shown in  Fig.~\ref{fig:FCavScan}~a) we observe a narrow peak (blue) centered at the expected heralding photon frequency. This peak results from the symmetric collective excitation while the broad peak (gray) centered at the same frequency comes from asymmetric excitations due to insufficient motional averaging. Furthermore we have a contribution from leakage (yellow) of the write excitation light centered at $\Delta_\text{FC} = - \nu_\mathrm{L}$ and negligible background counts. We observe good agreement between the data (circles) and a fit (line)
consisting of the sum of the narrow, broad, leakage and background contributions, which is given in the same order 
\begin{equation}
S_\text{W}(\Delta_\text{FC}) = a_\text{narr} \mathcal{L}_{1}(\Delta_\text{FC},0) \mathcal{L}_{2}(\Delta_\text{FC},0)
+ a_\text{broad} \mathcal{L}_\text{broad}(\Delta_\text{FC},0)  
+ a_\text{lkg} \mathcal{L}_{1}(\Delta_\text{FC},- \nu_\text{L}) \mathcal{L}_{2}(\Delta_\text{FC},- \nu_\text{L})
+ a_\text{bg}
\end{equation}
where $\mathcal{L}_i(\Delta_\text{FC},\Delta_0)$ is a Lorentzian centered at $\Delta_0$ with unity peak value
and $i$ indexes the linewidth of the according filter cavities ($1$, $2$) or the Lorentzian lineshape expected from the asymmetric excitations \cite{Borregaard2016} (labelled "broad") and $a_j$ is the amplitude of the respective contribution. 
The filter cavities are optimized for high transmission of the scattered photons, therefore the measured width of the narrow peak is mainly limited by the width of our filter cavities. We can determine the write efficiency $\eta_{W}$, that is the probability to create a symmetric excitation conditioned on the detection of a scattered photon, as the ratio $a_\text{narr}/(a_\text{narr}+a_\text{broad})$ resulting in $\eta_\text{W} = (82\pm1)\%$. This value is improved by $20\%$ compared to previously reported \cite{Zugenmaier2018} thanks to an increased excitation beam waist in the cell.

For the read process shown in  Fig.~\ref{fig:FCavScan} b) we observe the unconditional retrieval counts (circles) above the noise level (crosses) which is measured in the absence of a preceding write pulse.
Besides the desired narrowband retrieval of the stored excitation (light red) we observe a narrowband noise contribution (dark red), a broadband contribution (gray), leakage from read excitation (yellow) that is here centered at $+\nu_\mathrm{L}$ and negligible background counts.
Imperfect initial pumping and decay over the pulse durations leads to excess atomic population in the symmetric and asymmetric modes of the state $\ket{4,3}$ resulting in the observed narrowband and broadband noise, respectively.
We fit the expression
\begin{equation}
S^i_\text{R}(\Delta_\text{FC}) = b^i_\text{narr} \mathcal{L}_{1}(\Delta_\text{FC},0) \mathcal{L}_{2}(\Delta_\text{FC},0)
+ b_\text{broad} \mathcal{L}_\text{broad}(\Delta_\text{FC},0)  
+ b_\text{lkg} \mathcal{L}_{1}(\Delta_\text{FC},+ \nu_\text{L}) \mathcal{L}_{2}(\Delta_\text{FC},+ \nu_\text{L})
+ b_\text{bg} 
\end{equation}
simultaneous to the two data sets $i\in\{W,NW\}$, i.e. with and without preceding write pulse. Only the narrowband amplitudes $b^i_\text{narr}$ are specific to the data sets and the rest are common fit parameters.

\section{Modelling correlations}
\label{app:CorrModelling}
In order to establish analytical expressions for the photon correlations in our experiment in the presence of noise, we describe our system in the framework of probability generating functions. The probability generating function $G_X(s)$ of a process $X$ with outcomes $x$ is defined as 
\begin{equation}
   G_X(s)=\mathbb{E}(s^x)=\sum_{x=0}^{\infty} \mathbb{P}(X=x) s^x.
\end{equation}
We can derive the $k$'th factorial moment of the distribution
\begin{equation}
    \mathbb{E}\{X(X-1)...(X-k+1)\} = \left. \frac{d^k}{ds^k} G_X (s) \right|_{s=1}
\end{equation}
and express correlations as 
\begin{equation}
    G_{X,Y}(s,t) = \mathbb{E}(s^x t^y) = \sum_{x=0}^{\infty}\sum_{y=0}^{\infty} \mathbb{P}(X=x,Y=y) s^x t^y. 
\end{equation} 
The system is modelled by a two-mode squeezed state which is described by a thermal distribution of photons pairs in write ($X$) and read ($Y$) with the probability generating function
\begin{equation}
    G_{X,Y}(s,t) = \frac{1}{1+\mu(1-st)}
\end{equation}
where $\mu$ is the mean number of excitations. 
Then we add noise as independent processes for write ($W=X+A$) and read ($R=Y+B$) such that
\begin{equation}
    G_{W,R}(s,t) = G_{X,Y}(s,t)G_{A}(s)G_{B}(t)
\end{equation}
where the noise has mean numbers $\lambda_A$ and $\lambda_B$ respectively. Note, that these are the detected mean numbers. The actual form of $G_A,G_B$ is irrelevant for the derivation where only the mean values and 2$^{\text{nd}}$-order correlations are needed. 
The limited detection efficiency during write $\eta_X$ and read $\eta_Y$ is accounted for by substituting $s\rightarrow 1+\eta_X(s-1)$ and similarly for $t$. The detection efficiency includes all losses from generation of a photon to a detection event. 
For the read step this includes the intrinsic retrieval efficiency, that is the probability to generate a photon conditioned on a single collective symmetric excitation.

The probability generating functions  yield an expression for the cross-correlation
\begin{equation}
    g^{(2)}_{WR} = \frac{\mathbb{E}(WR)}{\mathbb{E}(W)\mathbb{E}(R)}
    =1+\frac{\eta_X\eta_Y(\mu^2+\mu)}{\eta_X\eta_Y\mu^2 + \mu(\eta_Y\lambda_A + \eta_X\lambda_B) + \lambda_A \lambda_B}.
\end{equation}
The conditional auto-correlation is found by first introducing the generator for the conditional read
\begin{equation}
G_{R|W=1}(t) = \frac{\left. \frac{d}{ds} G_{W,R} (s,t) \right|_{s=0}}{\left. \frac{d}{ds} G_{W,R} (s,t) \right|_{s=0,t=1}} \label{eq:autocorr}
\end{equation}
and recalling the definition of the 2$^{\text{nd}}$-order autocorrelation function
\begin{equation}
    g^{(2)}_{RR|W=1} = \frac{\mathbb{E}(R(R-1))}{\mathbb{E}(R)^2} = \frac{\eta_Y^2 \tilde{\mu}^2 g^{(2)}_{YY|W=1} + \lambda_B^2 g^{(2)}_{BB} + 2\eta_Y \tilde{\mu}\lambda_B}{\eta_Y^2 \tilde{\mu}^2 + \lambda_B^2 + 2\eta_Y \tilde{\mu}\lambda_B}
\end{equation}
where we have introduced the conditional mean excitation number $\tilde{\mu}=\mathbb{E}(Y|W=1)$, the auto-correlation of the read noise $g^{(2)}_{BB}$ and used 
\begin{equation}
    g^{(2)}_{YY|W=1}(\mu,\eta_X,\lambda_A) =
 \frac{-(2 (\eta_X - 1) (\lambda_A + \eta_X \mu + \lambda_A \eta_X \mu) (\lambda_A + 2 \eta_X - \lambda_A \eta_X + 3 \eta_X \mu - \eta_X^2 \mu + \lambda_A \eta_X \mu - \lambda_A \eta_X^2 \mu))}{(\lambda_A + \eta_X - \lambda_A \eta_X + 2 \eta_X \mu - \eta_X^2 \mu + \lambda_A \eta_X \mu - \lambda_A \eta_X^2 \mu)^2}.
\end{equation}
Furthermore, we can calculate the retrieval efficiency $\eta_R = \eta_Y \tilde{\mu}$ and the mean detected photon number in the read process $\left<n_\textrm{R}\right> = \mathbb{E}(R)$.

The above expressions depend in total on five parameters that are determined via calibration measurements.
To complete the model for different write excitation powers, we need to add a sixth parameter, that describes how the noise during the write excitation pulse $\lambda_A$ changes with the write power.
\begin{figure}[H]
    \centering
    \includegraphics[width=160mm]{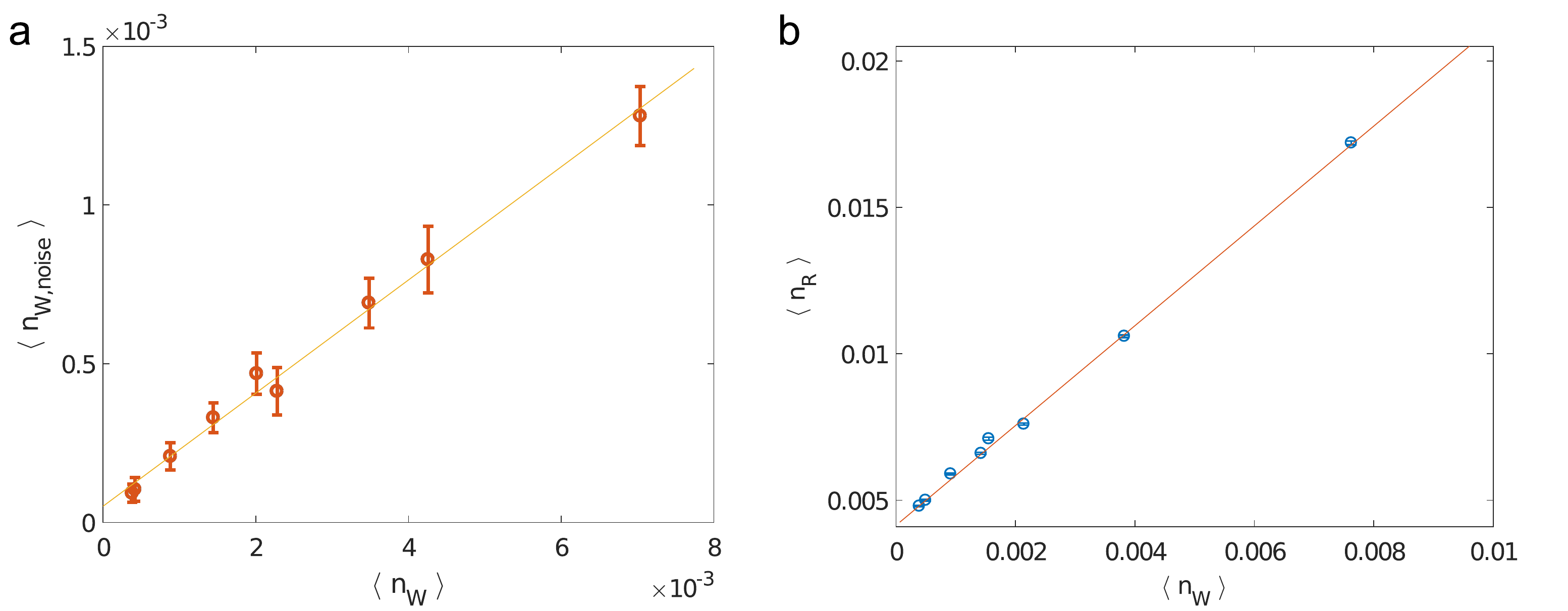}
    \caption{\textbf{a)} Mean number of noise counts per pulse measured during the write detection window (red circles) plotted versus the mean number of write counts with Poissonian standard deviation. Linear fit shown as yellow line.
    \textbf{b)} Mean number of counts per pulse measured during the read detection window (blue circles) plotted versus the mean number of write counts. Error bars are standard deviation assuming Poissonian distribution of $\langle n_R \rangle$. The fitted model is shown as the red line.
    }
    \label{fig:meanR}
\end{figure}%
We determine the write noise from the spectral analysis (Fig.\ \ref{fig:FCavScan}) as the on-resonance value when subtracting the narrowband peak.
We observe, that the write noise scales linearly with the mean number of write counts 
(Fig.\ \ref{fig:meanR} a). Fitting yields the offset background noise level and the slope that are then used as input parameters for the model.
We observe that the read noise level $\lambda_B$ and the auto-correlation of this noise $g^{(2)}_{BB}$ are constant in the write power and determine their values by averaging the measurement results taken without a preceding write excitation pulse. This leaves the two detection efficiencies as free parameters. 

Those parameters are determined by simultaneously fitting \gWRcrosscorr{}, $\eta_\textrm{R}$ and $\left<n_\textrm{R}\right>$. We observe good agreement of the experimental data with the fitted model shown as the curves in the respective figure in the main text and in the supplementary Fig.~\ref{fig:meanR} \textbf{b}).

We obtain the the best fit values for the detection efficiencies of $\eta_X = (2.9\pm0.1)\%$ and $\eta_Y = (6.0\pm0.2)\%$. 

\section{Retrieval efficiency vs. delay}
\label{app:RetEffvsTauD}
The retrieval efficiency $\eta_\textrm{R}$ determines how well we can retrieve a stored excitation conditioned on detection of a preceding heralding write single photon, and is defined as 
\begin{equation}
    \eta_\textrm{R} \quad = \quad \braket{n_{\textrm{R|W=1}}} - \braket{n_{\textrm{noise}}}
\end{equation}
where $\braket{n_{\textrm{R|W=1}}} $ is the mean number of read detection events conditioned on a single preceding write heralding click, and $\braket{n_{\textrm{noise}}}$ is the mean number of read detection events obtained without write pulses.
In this way, the retrieval efficiency is intrinsically corrected for noise.

\begin{figure}[H]
    \centering
    \includegraphics[width= 0.7\textwidth]{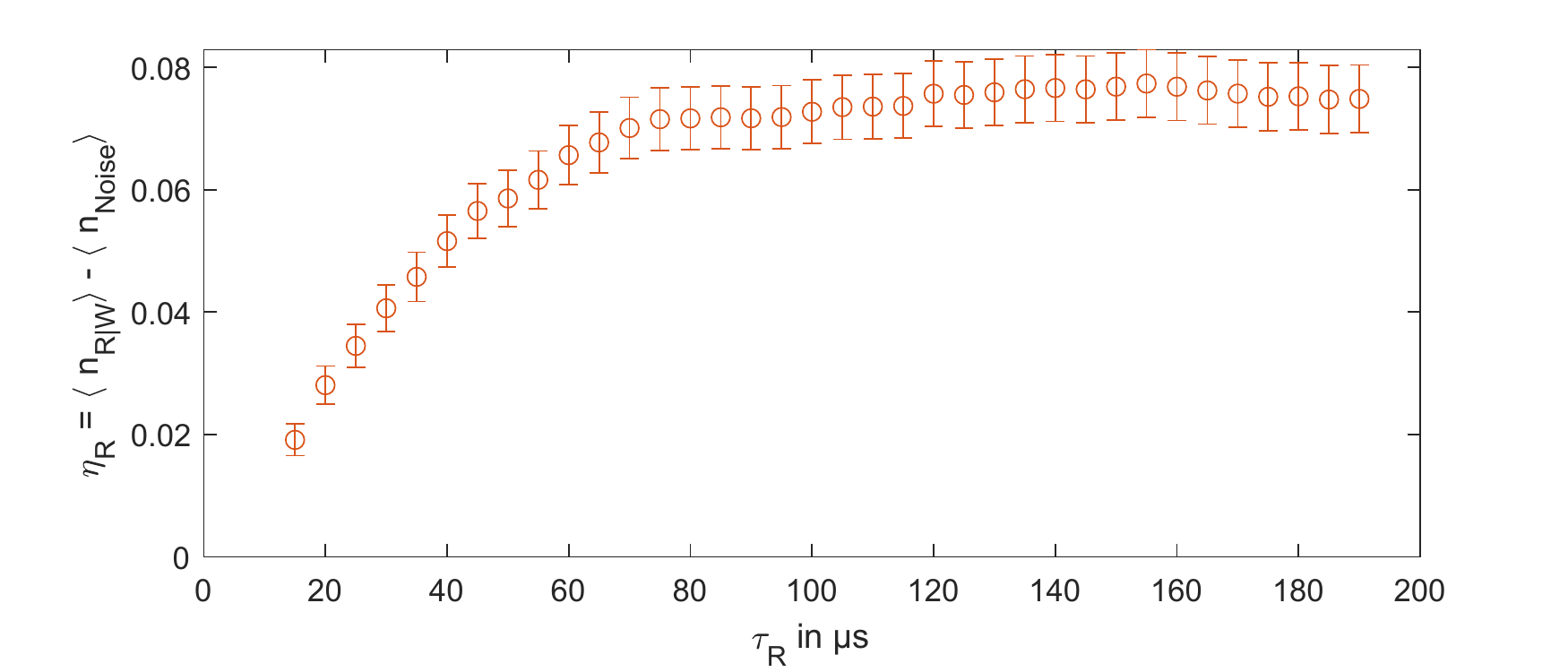}
    \caption{\textbf{Retrieval efficiency $\eta_\textrm{R}$ vs. integration time $\tau_\textrm{R}$.} 
    Error bars are standard deviation assuming Poissonian distribution of count rates.
    }
    \label{fig:RetEff_vs_tauR}
\end{figure}

In our experiment, the total read duration is \readpulse, while for the analysis we choose to integrate only over the first $\SI{40}{\micro\second}$. The reason is, that there is a trade-off between signal-to-noise ratio and retrieval efficiency. From Fig.\ \ref{fig:histograms_cond_uncond_readpulses} it is apparent, that the SNR decreases for longer read pulse integration window $\tau_\text{R}$. At the same time however, the retrieval efficiency grows with $\tau_\text{R}$, as can be seen in Fig.\ \ref{fig:RetEff_vs_tauR}. 
Note that the data points in Fig.\ \ref{fig:RetEff_vs_tauR} are not statistically independent since only the analysis parameter $\tau_\text{R}$ is varied.

By choosing $\tau_\text{R}=$\tauR, we achieve good SNR, while still maintaining high retrieval efficiency. Furthermore, the estimated intrinsic retrieval efficiency $\eta^*_\text{R} = (70\pm8)\%$ for $\tau_\text{R}=$\tauR indicates that we observe close to 100\% of the full read pulse.

\begin{figure}[H]
    \centering
    \includegraphics[width= 0.7\textwidth]{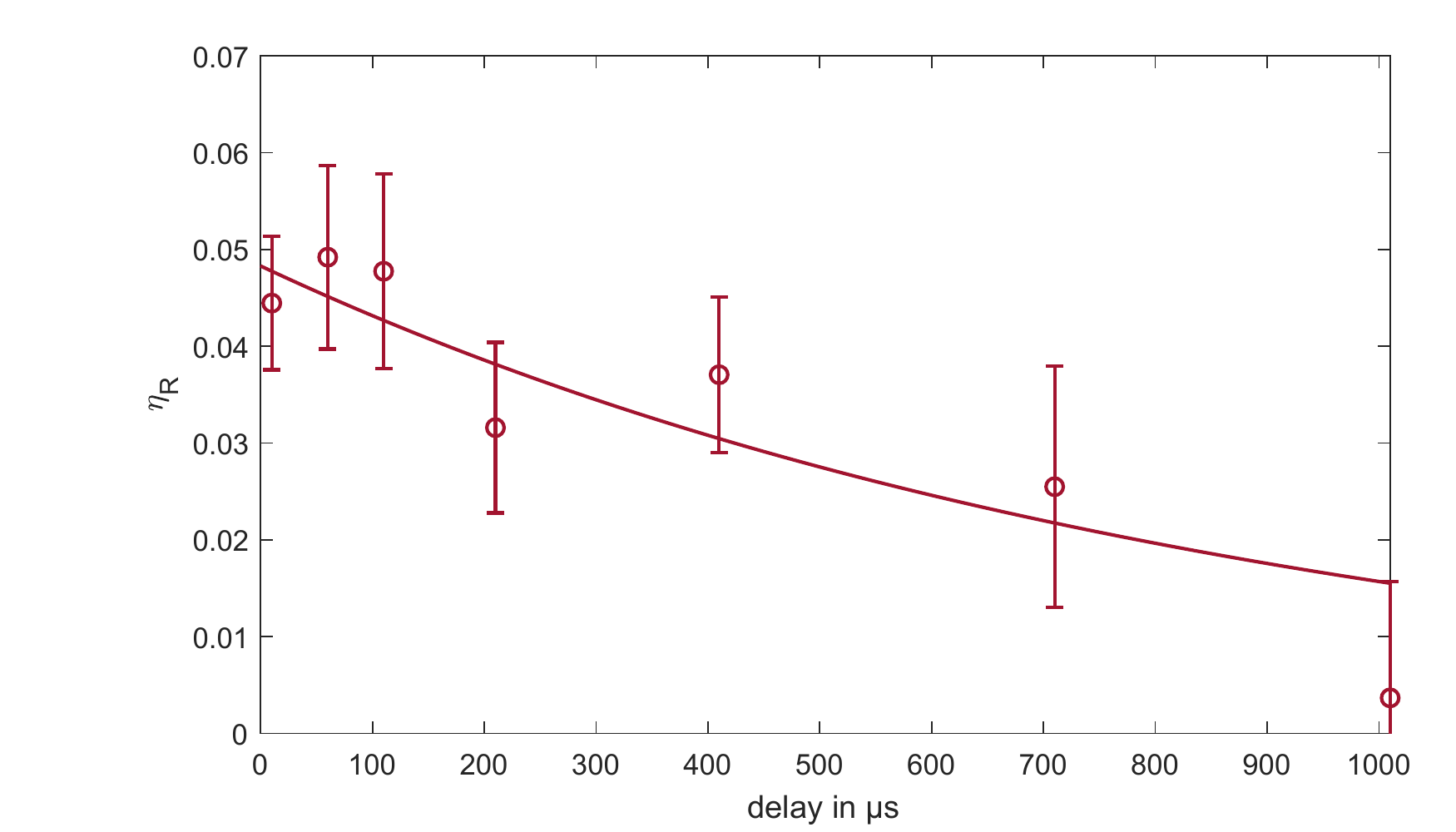}
    \caption{\textbf{Retrieval efficiency $\eta_\textrm{R}$ vs. delay time $\tau_\text D:$} Shown is the retrieval efficiency for various delay times and a fixed readout integration duration of $\tau_\textrm{R} = \SI{40}{\micro\second}$. The uncertainty on the retrieval efficiency is calculated using Poissonian errors. An exponential function is fit to obtain the intrinsic memory time. 
    }
    \label{fig:RetEff_vs_tauD}
\end{figure}

Since the retrieval efficiency is corrected for noise by definition, it can be used to determine the lifetime of the collective excitation. To this end, the retrieval efficiency is determined for various delays $\tau_\textrm{D}$ and the resulting values are fit to an exponential fit model:
\begin{equation}
    \eta_\textrm{R}(\tau_\textrm{D}) \quad = \quad B\cdot\exp\left({-\tau_\textrm{D}/\tau_{\eta_\text{R}}}\right).
\end{equation}
From the resulting fit in Fig.\ \ref{fig:RetEff_vs_tauD}, the memory time for the retrieval efficiency is determined as the time, during which the retrieval efficiency is reduced to its $1/e$-value, \memorytimeretrieval.

\begin{thebibliography}{10}
	\expandafter\ifx\csname url\endcsname\relax
	\def\url#1{\texttt{#1}}\fi
	\expandafter\ifx\csname urlprefix\endcsname\relax\def\urlprefix{URL }\fi
	\providecommand{\bibinfo}[2]{#2}
	\providecommand{\eprint}[2][]{\url{#2}}
	
	\bibitem{Aharonovich2016}
	\bibinfo{author}{Aharonovich, I.}, \bibinfo{author}{Englund, D.} \&
	\bibinfo{author}{Toth, M.}
	\newblock \bibinfo{title}{{Solid-state single-photon emitters}}.
	\newblock \emph{\bibinfo{journal}{Nature Photonics}}
	\textbf{\bibinfo{volume}{10}}, \bibinfo{pages}{631--641}
	(\bibinfo{year}{2016}).
	
	\bibitem{Awschalom2018}
	\bibinfo{author}{Awschalom, D.~D.}, \bibinfo{author}{Hanson, R.},
	\bibinfo{author}{Wrachtrup, J.} \& \bibinfo{author}{Zhou, B.~B.}
	\newblock \bibinfo{title}{{Quantum technologies with optically interfaced
			solid-state spins}}.
	\newblock \emph{\bibinfo{journal}{Nature Photonics}}
	\textbf{\bibinfo{volume}{12}}, \bibinfo{pages}{516--527}
	(\bibinfo{year}{2018}).
	
	\bibitem{Wan2020}
	\bibinfo{author}{Wan, N.~H.} \emph{et~al.}
	\newblock \bibinfo{title}{{Large-scale integration of artificial atoms in
			hybrid photonic circuits}}.
	\newblock \emph{\bibinfo{journal}{Nature}} \textbf{\bibinfo{volume}{583}},
	\bibinfo{pages}{226--231} (\bibinfo{year}{2020}).
	
	\bibitem{Ding2016}
	\bibinfo{author}{Ding, X.} \emph{et~al.}
	\newblock \bibinfo{title}{{On-Demand Single Photons with High Extraction
			Efficiency and Near-Unity Indistinguishability from a Resonantly Driven
			Quantum Dot in a Micropillar}}.
	\newblock \emph{\bibinfo{journal}{Physical Review Letters}}
	\textbf{\bibinfo{volume}{116}}, \bibinfo{pages}{020401}
	(\bibinfo{year}{2016}).
	
	\bibitem{Kirsanske2017}
	\bibinfo{author}{Kir{\v{s}}anskė, G.} \emph{et~al.}
	\newblock \bibinfo{title}{{Indistinguishable and efficient single photons from
			a quantum dot in a planar nanobeam waveguide}}.
	\newblock \emph{\bibinfo{journal}{Physical Review B}}
	\textbf{\bibinfo{volume}{96}}, \bibinfo{pages}{165306}
	(\bibinfo{year}{2017}).
	
	\bibitem{Laplane2017}
	\bibinfo{author}{Laplane, C.}, \bibinfo{author}{Jobez, P.},
	\bibinfo{author}{Etesse, J.}, \bibinfo{author}{Gisin, N.} \&
	\bibinfo{author}{Afzelius, M.}
	\newblock \bibinfo{title}{{Multimode and Long-Lived Quantum Correlations
			between Photons and Spins in a Crystal}}.
	\newblock \emph{\bibinfo{journal}{Physical Review Letters}}
	\textbf{\bibinfo{volume}{118}}, \bibinfo{pages}{210501}
	(\bibinfo{year}{2017}).
	
	\bibitem{Kutluer2017}
	\bibinfo{author}{Kutluer, K.}, \bibinfo{author}{Mazzera, M.} \&
	\bibinfo{author}{{De Riedmatten}, H.}
	\newblock \bibinfo{title}{{Solid-State Source of Nonclassical Photon Pairs with
			Embedded Multimode Quantum Memory}}.
	\newblock \emph{\bibinfo{journal}{Physical Review Letters}}
	\textbf{\bibinfo{volume}{118}}, \bibinfo{pages}{210502}
	(\bibinfo{year}{2017}).
	
	\bibitem{Lukishova2019}
	\bibinfo{author}{Lukishova, S.~G.} \& \bibinfo{author}{Bissell, L.~J.}
	\newblock \emph{\bibinfo{title}{Nanophotonic Advances for Room-Temperature
			Single-Photon Sources}}, \bibinfo{pages}{103--178}
	(\bibinfo{publisher}{Springer International Publishing},
	\bibinfo{address}{Cham}, \bibinfo{year}{2019}).
	
	\bibitem{Mucke2013}
	\bibinfo{author}{M{\"{u}}cke, M.} \emph{et~al.}
	\newblock \bibinfo{title}{{Generation of single photons from an atom-cavity
			system}}.
	\newblock \emph{\bibinfo{journal}{Physical Review A}}
	\textbf{\bibinfo{volume}{87}}, \bibinfo{pages}{063805}
	(\bibinfo{year}{2013}).
	
	\bibitem{Radnaev2010}
	\bibinfo{author}{Radnaev, A.~G.} \emph{et~al.}
	\newblock \bibinfo{title}{{A quantum memory with telecom-wavelength
			conversion}}.
	\newblock \emph{\bibinfo{journal}{Nature Physics}}
	\textbf{\bibinfo{volume}{6}}, \bibinfo{pages}{894--899}
	(\bibinfo{year}{2010}).
	
	\bibitem{Corzo2019}
	\bibinfo{author}{Corzo, N.~V.} \emph{et~al.}
	\newblock \bibinfo{title}{{Waveguide-coupled single collective excitation of
			atomic arrays}}.
	\newblock \emph{\bibinfo{journal}{Nature}} \textbf{\bibinfo{volume}{566}},
	\bibinfo{pages}{359--362} (\bibinfo{year}{2019}).
	
	\bibitem{Yang2016}
	\bibinfo{author}{Yang, S.~J.}, \bibinfo{author}{Wang, X.~J.},
	\bibinfo{author}{Bao, X.~H.} \& \bibinfo{author}{Pan, J.~W.}
	\newblock \bibinfo{title}{{An efficient quantum light-matter interface with
			sub-second lifetime}}.
	\newblock \emph{\bibinfo{journal}{Nature Photonics}}
	\textbf{\bibinfo{volume}{10}}, \bibinfo{pages}{381--384}
	(\bibinfo{year}{2016}).
	
	\bibitem{Bimbard2014}
	\bibinfo{author}{Bimbard, E.} \emph{et~al.}
	\newblock \bibinfo{title}{{Homodyne tomography of a single photon retrieved on
			demand from a cavity-enhanced cold atom memory}}.
	\newblock \emph{\bibinfo{journal}{Physical Review Letters}}
	\textbf{\bibinfo{volume}{112}}, \bibinfo{pages}{033601}
	(\bibinfo{year}{2014}).
	
	\bibitem{Shu2016}
	\bibinfo{author}{Shu, C.} \emph{et~al.}
	\newblock \bibinfo{title}{{Subnatural-linewidth biphotons from a
			Doppler-broadened hot atomic vapour cell}}.
	\newblock \emph{\bibinfo{journal}{Nature Communications}}
	\textbf{\bibinfo{volume}{7}}, \bibinfo{pages}{12783} (\bibinfo{year}{2016}).
	
	\bibitem{Eisaman2005}
	\bibinfo{author}{Eisaman, M.~D.} \emph{et~al.}
	\newblock \bibinfo{title}{{Electromagnetically induced transparency with
			tunable single-photon pulses}}.
	\newblock \emph{\bibinfo{journal}{Nature}} \textbf{\bibinfo{volume}{438}},
	\bibinfo{pages}{837--841} (\bibinfo{year}{2005}).
	
	\bibitem{Ripka2018}
	\bibinfo{author}{Ripka, F.}, \bibinfo{author}{K{\"{u}}bler, H.},
	\bibinfo{author}{L{\"{o}}w, R.} \& \bibinfo{author}{Pfau, T.}
	\newblock \bibinfo{title}{{A room-temperature single-photon source based on
			strongly interacting Rydberg atoms}}.
	\newblock \emph{\bibinfo{journal}{Science}} \textbf{\bibinfo{volume}{362}},
	\bibinfo{pages}{446--449} (\bibinfo{year}{2018}).
	
	\bibitem{Finkelstein2018}
	\bibinfo{author}{Finkelstein, R.}, \bibinfo{author}{Poem, E.},
	\bibinfo{author}{Michel, O.}, \bibinfo{author}{Lahad, O.} \&
	\bibinfo{author}{Firstenberg, O.}
	\newblock \bibinfo{title}{{Fast, noise-free memory for photon synchronization
			at room temperature}}.
	\newblock \emph{\bibinfo{journal}{Science Advances}}
	\textbf{\bibinfo{volume}{4}}, \bibinfo{pages}{eaap8598}
	(\bibinfo{year}{2018}).
	
	\bibitem{Kaczmarek2018}
	\bibinfo{author}{Kaczmarek, K.~T.} \emph{et~al.}
	\newblock \bibinfo{title}{{High-speed noise-free optical quantum memory}}.
	\newblock \emph{\bibinfo{journal}{Physical Review A}}
	\textbf{\bibinfo{volume}{97}}, \bibinfo{pages}{042316}
	(\bibinfo{year}{2018}).
	
	\bibitem{Duan2001}
	\bibinfo{author}{Duan, L.~M.}, \bibinfo{author}{Lukin, M.~D.},
	\bibinfo{author}{Cirac, J.~I.} \& \bibinfo{author}{Zoller, P.}
	\newblock \bibinfo{title}{{Long-distance quantum communication with atomic
			ensembles and linear optics}}.
	\newblock \emph{\bibinfo{journal}{Nature}} \textbf{\bibinfo{volume}{414}},
	\bibinfo{pages}{413--418} (\bibinfo{year}{2001}).
	
	\bibitem{Nunn2013}
	\bibinfo{author}{Nunn, J.} \emph{et~al.}
	\newblock \bibinfo{title}{{Enhancing multiphoton rates with quantum memories}}.
	\newblock \emph{\bibinfo{journal}{Physical Review Letters}}
	\textbf{\bibinfo{volume}{110}}, \bibinfo{pages}{133601}
	(\bibinfo{year}{2013}).
	
	\bibitem{Sangouard2011}
	\bibinfo{author}{Sangouard, N.}, \bibinfo{author}{Simon, C.},
	\bibinfo{author}{{De Riedmatten}, H.} \& \bibinfo{author}{Gisin, N.}
	\newblock \bibinfo{title}{{Quantum repeaters based on atomic ensembles and
			linear optics}}.
	\newblock \emph{\bibinfo{journal}{Reviews of Modern Physics}}
	\textbf{\bibinfo{volume}{83}}, \bibinfo{pages}{33--80}
	(\bibinfo{year}{2011}).
	
	\bibitem{Dou2018}
	\bibinfo{author}{Dou, J.~P.} \emph{et~al.}
	\newblock \bibinfo{title}{{A broadband DLCZ quantum memory in room-temperature
			atoms}}.
	\newblock \emph{\bibinfo{journal}{Communications Physics}}
	\textbf{\bibinfo{volume}{1}}, \bibinfo{pages}{55} (\bibinfo{year}{2018}).
	
	\bibitem{Pang2020}
	\bibinfo{author}{Pang, X.~L.} \emph{et~al.}
	\newblock \bibinfo{title}{{A hybrid quantum memory–enabled network at room
			temperature}}.
	\newblock \emph{\bibinfo{journal}{Science Advances}}
	\textbf{\bibinfo{volume}{6}}, \bibinfo{pages}{eaax1425}
	(\bibinfo{year}{2020}).
	
	\bibitem{Bashkansky2012}
	\bibinfo{author}{Bashkansky, M.}, \bibinfo{author}{Fatemi, F.~K.} \&
	\bibinfo{author}{Vurgaftman, I.}
	\newblock \bibinfo{title}{{Quantum memory in warm rubidium vapor with buffer
			gas}}.
	\newblock \emph{\bibinfo{journal}{Optics Letters}}
	\textbf{\bibinfo{volume}{37}}, \bibinfo{pages}{142--144}
	(\bibinfo{year}{2012}).
	
	\bibitem{Namazi2017}
	\bibinfo{author}{Namazi, M.}, \bibinfo{author}{Kupchak, C.},
	\bibinfo{author}{Jordaan, B.}, \bibinfo{author}{Shahrokhshahi, R.} \&
	\bibinfo{author}{Figueroa, E.}
	\newblock \bibinfo{title}{{Ultralow-Noise Room-Temperature Quantum Memory for
			Polarization Qubits}}.
	\newblock \emph{\bibinfo{journal}{Physical Review Applied}}
	\textbf{\bibinfo{volume}{8}}, \bibinfo{pages}{034023} (\bibinfo{year}{2017}).
	
	\bibitem{Reim2011}
	\bibinfo{author}{Reim, K.~F.} \emph{et~al.}
	\newblock \bibinfo{title}{{Single-Photon-Level Quantum Memory at Room
			Temperature}}.
	\newblock \emph{\bibinfo{journal}{Physical Review Letters}}
	\textbf{\bibinfo{volume}{107}}, \bibinfo{pages}{053603}
	(\bibinfo{year}{2011}).
	
	\bibitem{Hosseini2011}
	\bibinfo{author}{Hosseini, M.}, \bibinfo{author}{Campbell, G.},
	\bibinfo{author}{Sparkes, B.~M.}, \bibinfo{author}{Lam, P.~K.} \&
	\bibinfo{author}{Buchler, B.~C.}
	\newblock \bibinfo{title}{{Unconditional room-temperature quantum memory}}.
	\newblock \emph{\bibinfo{journal}{Nature Physics}}
	\textbf{\bibinfo{volume}{7}}, \bibinfo{pages}{794--798}
	(\bibinfo{year}{2011}).
	
	\bibitem{Zugenmaier2018}
	\bibinfo{author}{Zugenmaier, M.}, \bibinfo{author}{Dideriksen, K.~B.},
	\bibinfo{author}{S{\o}rensen, A.~S.}, \bibinfo{author}{Albrecht, B.} \&
	\bibinfo{author}{Polzik, E.~S.}
	\newblock \bibinfo{title}{{Long-lived non-classical correlations towards
			quantum communication at room temperature}}.
	\newblock \emph{\bibinfo{journal}{Communications Physics}}
	\textbf{\bibinfo{volume}{1}}, \bibinfo{pages}{76} (\bibinfo{year}{2018}).
	
	\bibitem{Borregaard2016}
	\bibinfo{author}{Borregaard, J.} \emph{et~al.}
	\newblock \bibinfo{title}{{Scalable photonic network architecture based on
			motional averaging in room temperature gas}}.
	\newblock \emph{\bibinfo{journal}{Nature Communications}}
	\textbf{\bibinfo{volume}{7}}, \bibinfo{pages}{11356} (\bibinfo{year}{2016}).
	
	\bibitem{Balabas2010}
	\bibinfo{author}{Balabas, M.~V.}, \bibinfo{author}{Karaulanov, T.},
	\bibinfo{author}{Ledbetter, M.~P.} \& \bibinfo{author}{Budker, D.}
	\newblock \bibinfo{title}{{Polarized alkali-metal vapor with minute-long
			transverse spin-relaxation time}}.
	\newblock \emph{\bibinfo{journal}{Physical Review Letters}}
	\textbf{\bibinfo{volume}{105}}, \bibinfo{pages}{070801}
	(\bibinfo{year}{2010}).
	
	\bibitem{Dicke1954}
	\bibinfo{author}{Dicke, R.~H.}
	\newblock \bibinfo{title}{{Coherence in Spontaneous Radiation Processes}}.
	\newblock \emph{\bibinfo{journal}{Physical Review}}
	\textbf{\bibinfo{volume}{93}}, \bibinfo{pages}{99--110}
	(\bibinfo{year}{1954}).
	
	\bibitem{Michelberger2015}
	\bibinfo{author}{Michelberger, P.~S.} \emph{et~al.}
	\newblock \bibinfo{title}{{Interfacing GHz-bandwidth heralded single photons
			with a warm vapour Raman memory}}.
	\newblock \emph{\bibinfo{journal}{New Journal of Physics}}
	\textbf{\bibinfo{volume}{17}}, \bibinfo{pages}{043006}
	(\bibinfo{year}{2015}).
	
	\bibitem{Walther2007}
	\bibinfo{author}{Walther, P.} \emph{et~al.}
	\newblock \bibinfo{title}{{Generation of narrow-bandwidth single photons using
			electromagnetically induced transparency in atomic ensembles}}.
	\newblock \emph{\bibinfo{journal}{International Journal of Quantum
			Information}} \textbf{\bibinfo{volume}{5}}, \bibinfo{pages}{51--62}
	(\bibinfo{year}{2007}).
	
	\bibitem{Saunders2016}
	\bibinfo{author}{Saunders, D.~J.} \emph{et~al.}
	\newblock \bibinfo{title}{Cavity-enhanced room-temperature broadband raman
		memory}.
	\newblock \emph{\bibinfo{journal}{Phys. Rev. Lett.}}
	\textbf{\bibinfo{volume}{116}}, \bibinfo{pages}{090501}
	(\bibinfo{year}{2016}).
	
	\bibitem{Thomas2019}
	\bibinfo{author}{Thomas, S.~E.} \emph{et~al.}
	\newblock \bibinfo{title}{{Raman quantum memory with built-in suppression of
			four-wave-mixing noise}}.
	\newblock \emph{\bibinfo{journal}{Physical Review A}}
	\textbf{\bibinfo{volume}{100}}, \bibinfo{pages}{33801}
	(\bibinfo{year}{2019}).
	
	\bibitem{Romanov2016}
	\bibinfo{author}{Romanov, G.}, \bibinfo{author}{O'Brien, C.} \&
	\bibinfo{author}{Novikova, I.}
	\newblock \bibinfo{title}{{Suppression of the four-wave mixing amplification
			via Raman absorption}}.
	\newblock \emph{\bibinfo{journal}{Journal of Modern Optics}}
	\textbf{\bibinfo{volume}{63}}, \bibinfo{pages}{2048--2057}
	(\bibinfo{year}{2016}).
	
	\bibitem{Zhang2014}
	\bibinfo{author}{Zhang, K.} \emph{et~al.}
	\newblock \bibinfo{title}{Suppression of the four-wave-mixing background noise
		in a quantum memory retrieval process by channel blocking}.
	\newblock \emph{\bibinfo{journal}{Phys. Rev. A}} \textbf{\bibinfo{volume}{90}},
	\bibinfo{pages}{033823} (\bibinfo{year}{2014}).
	
	\bibitem{Vurgaftman2013}
	\bibinfo{author}{Vurgaftman, I.} \& \bibinfo{author}{Bashkansky, M.}
	\newblock \bibinfo{title}{{Suppressing four-wave mixing in warm-atomic-vapor
			quantum memory}}.
	\newblock \emph{\bibinfo{journal}{Physical Review A}}
	\textbf{\bibinfo{volume}{87}}, \bibinfo{pages}{063836}
	(\bibinfo{year}{2013}).
	
	\bibitem{Clauser1974}
	\bibinfo{author}{Clauser, J.~F.}
	\newblock \bibinfo{title}{{Experimental distinction between the quantum and
			classical field-theoretic predictions for the photoelectric effect}}.
	\newblock \emph{\bibinfo{journal}{Physical Review D}}
	\textbf{\bibinfo{volume}{9}}, \bibinfo{pages}{853--860}
	(\bibinfo{year}{1974}).
	
	\bibitem{Wallucks2020}
	\bibinfo{author}{Wallucks, A.}, \bibinfo{author}{Marinkovi{\'{c}}, I.},
	\bibinfo{author}{Hensen, B.}, \bibinfo{author}{Stockill, R.} \&
	\bibinfo{author}{Gr{\"{o}}blacher, S.}
	\newblock \bibinfo{title}{{A quantum memory at telecom wavelengths}}.
	\newblock \emph{\bibinfo{journal}{Nature Physics}}
	\textbf{\bibinfo{volume}{16}}, \bibinfo{pages}{772--777}
	(\bibinfo{year}{2020}).
	
	\bibitem{Li2020}
	\bibinfo{author}{Li, H.} \emph{et~al.}
	\newblock \bibinfo{title}{{Heralding Quantum Entanglement between Two
			Room-Temperature Atomic Ensembles.}} \bibinfo{note}{Preprint at
		\url{http://arxiv.org/abs/2007.10948} (2020)}.
	
	\bibitem{Katz2018}
	\bibinfo{author}{Katz, O.} \& \bibinfo{author}{Firstenberg, O.}
	\newblock \bibinfo{title}{{Light storage for one second in room-temperature
			alkali vapor}}.
	\newblock \emph{\bibinfo{journal}{Nature Communications}}
	\textbf{\bibinfo{volume}{9}}, \bibinfo{pages}{2074} (\bibinfo{year}{2018}).
	
	\bibitem{Julsgaard2004}
	\bibinfo{author}{Julsgaard, B.}, \bibinfo{author}{Sherson, J.},
	\bibinfo{author}{S{\o}rensen, J.~L.} \& \bibinfo{author}{Polzik, E.~S.}
	\newblock \bibinfo{title}{{Characterizing the spin state of an atomic ensemble
			using the magneto-optical resonance method}}.
	\newblock \emph{\bibinfo{journal}{Journal of Optics B: Quantum and
			Semiclassical Optics}} \textbf{\bibinfo{volume}{6}}, \bibinfo{pages}{5--14}
	(\bibinfo{year}{2004}).
	
\end{thebibliography}
\end{document}